\documentclass[pra, twocolumn, showpacs, preprintnumbers, amsmath, amssymb, superscriptaddress, aps]{revtex4}
\usepackage{amssymb}
\usepackage{latexsym}

\usepackage{epsfig}

\usepackage{dcolumn}
\usepackage{amsmath}
\usepackage{amsfonts}
\usepackage{bm}

\newcommand{\be}{\begin{equation}}
\newcommand{\ee}{\end{equation}}
\newcommand{\bea}{\begin{eqnarray}}
\newcommand{\eea}{\end{eqnarray}}

\newcommand{\p}{\partial}
\newcommand{\s}{\sigma}

\newcommand{\la}{\langle}
\newcommand{\ra}{\rangle}
\newcommand{\rd}{\mbox{d}}
\newcommand{\ri}{\mbox{i}}
\newcommand{\re}{\mbox{e}}

\begin{document}
\title{A model description of the supersolid state in YBCO. }
\author{A.M. Tsvelik}
\affiliation{Department of Condensed Matter Physics and Materials Science, Brookhaven National Laboratory,
  Upton, NY 11973-5000, USA}
 \date{\today } \begin{abstract}  I employ a semiphenomenological model introduced in A. M. Tsvelik, A. V. Chubukov, Phys. Rev. Lett. {\bf 98}, 237001 (2007) to describe the state with co-existing superconductivity (SC) and charge density wave (CDW) recently discovered in YBCO.  The SC and the CDW order parameter fields are united in a single pseudospin and can be rotated into each other. It is suggested that disorder creates isolated pseudospins which become centers of inelastic scattering of electrons. It is suggested that this scattering is responsible for  the logarithmic upturn in the resistivity $\rho(T) \sim - \ln T$ observed at low doping. \end{abstract}


\maketitle
%
%

Recent x-ray measurements in clean YBа$_2$Cu$_3$O$_{6+x}$ \cite{xray},\cite{xray2} show that in a certain doping interval superconductivity in this material coexists either with Charge Density Wave (CDW) order or at least with strong CDW fluctuations.  The  CDW peak  emerges at temperatures below 150K and is not accompanied by  enhancement of the magnetic fluctuations.  According to \cite{hinkov}, an incommensurate SDW appears  at smaller dopings, closer to the ones where the  superconductivity disappears altogether.  Application of magnetic field suppresses the  superconductivity, but the CDW remains. Its presence should  lead to a reconstruction of the Fermi surface (FS) which has been invoked by Sebastian {\it et.al.} \cite{gil} as an explanation of  the results of de Haas van Alphen effect measurements \cite{gil},\cite{louis},\cite{louis2}. However, since the bulk of the FS is  already gapped at these temperatures, as is evident from the sharp drop in the integrated intensity of spin excitations occuring at $\sim$ 20 meV \cite{spingap}, the CDW  cannot be driven by such quasiparticle mechanisms as the FS nesting. It is reasonable therefore to suggest  that low energy physics of underdoped cuprates is dominated  by collective excitations and the  quasiparticles play a secondary role in determining their phase diagram (not in the transport where their role is important).

  This is the point of view adopted in this paper.  I consider a model which describes coexistence and competition of SC (superconductivity) and CDW in a state of preformed pairs where quasiparticles play a secondary role. At low temperatures the pairs may either localize (the CDW insulator) or exist in a supersolid state where charge density fluctuations coexist with superconductivity. The bulk of the bare FS is supposed to be gapped; there is a small quasiparticle FS whose area scales with doping.  
Within this model at zero temperature CDW+SC(supersolid) phase\cite{falicov} is separated from the pure CDW by a Quantum Critical Point (QCP). It would be logical to associate with it  the line of QCP's in $H-x$ plane ($H$ is magnetic field, $x$ is doping) observed in the conductivity measurements \cite{Hall},\cite{alloul} where  SC has been suppressed by magnetic field. However, the present x-ray data \cite{xray2} contradict this suggestion: the CDW peak disappears below $х$=0.09, well before the critical doping $х$=0.04.

  Universality of  Quantum Critical Point physics relieves one from concerns about microscopic mechanisms behind the adopted low energy effective theory giving a certain freedom in its choice. I choose a model capable to  describe a competition and coexistence between CDW and SC and  broadly compatible with the known cuprate physics.  

 The phenomenological model I use to extract the low energy properties are inspired by 1D physics, but can be justified by the DMFT results. These results as described in \cite{dmft} suggest that, as far as observables are concerned, {\it strongly interacting} 2D systems and  1D systems  are quite similar. First and foremost, the DMFT suggests that the pseudogap phenomenon is not  necessarily accompanied by any kind of order \cite{tremblay} which is similar  to dynamical generation of spectral gaps in 1D models. Second, there is a striking  similarity in detail between results of \cite{dmft},\cite{dmft1} and the physics of doped fermionic ladders as discussed, for instance, in \cite{tsvelik}. Such similarity has inspired now famous Yang-Rice-Zhang (YRZ) ansatz \cite{yrz} successfully employed as a phenomenological tool in the cuprate physics. In the latter case there is  also  a spin gap and singularities in the electronic self energy not accompanied by any obvious density wave order. The obvious difference though is that in 1D systems all these phenomena occur at any bare couplings due to the always present nesting, but  in 2D one needs strong interactions.


 Although I am  mostly interested in universal low energy physics, it would be desirable to use as a starting point more or less realistic lattice models. To illustrate the physics relevant to the QCP  I choose a model of crossed one-dimensional chains forming a grid introduced in \cite{chub}. The grid network is supposed to emulate the
interplay between CDW (or stripes) and superconductivity in terms of
preformed Cooper pairs, which appear naturally in 1D if the
interaction is attractive.
 Each chain contains electrons  with a strong local attraction:
 \bea
  && H_x = \sum_{{\bf r} = (a_0n_x,)}\Big\{ - t_x\Big[C^+_{\s}({\bf r} + {\bf x})C_{\s}({\bf r}) + H.c.\Big] -  \nonumber\\-
  && UN_{\uparrow}({\bf r})N_{\downarrow}({\bf r}) - \mu \sum_{{\s}}N_{\s}({\bf r})\Big\}, \label{Hubbard}
 \eea
where $N = \sum_{\s}C^+_{\s}C_{\s}$. The same Hamiltonian describes chains running parallel to $y$-axis. I denote its creation and annihilation operators $B^+,B$.  The electrons are allowed to tunnel
from the chains along $x$ to the chains along $y$ and viceversa at the
intersections. The Hamiltonian of crossed chains is 
\bea
&& H = \sum_{n_y} H_x[n_y] + \sum_{n_x} H_y[n_x] + \label{Model}\\
&& t_{\perp}\sum_{\s,n_x,n_y}[C^+_{\s}(a_0n_x,a_0n_y)B_{\s}(a_0Nn_x,a_0Nn_y) + H.c.],\nonumber
\eea
 where $a_0$ is a distance between parallel chains. I assume that $ t_{\perp} << t_x << U$. 

\begin{figure}[htp]
\centerline{\includegraphics[angle = 0,
width=1\columnwidth]{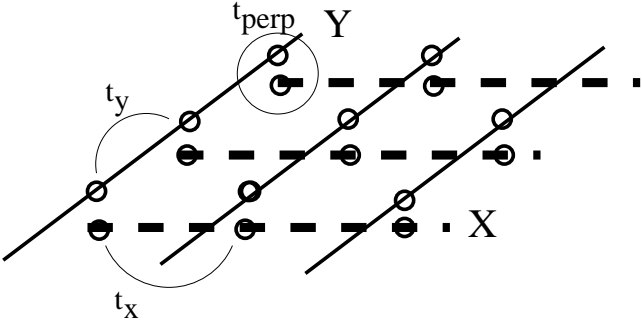}}
\caption{The grid-like lattice of model (\ref{Model}). The hollow circles mark lattice sites where $B,B^+$ and $C,C^+$ operators are defined. The solid and dashed lines mark tunneling pathes along $y$ and $x$ axes respectively.
}
\label{Fig1}
\end{figure}

 As I said, the sole purpose of model (\ref{Model})  is to surve as a seed for universal low energy physics which will be sensitive only to its gross features so that all nonrealistic ones will be "forgotten". To obtain this low energy description I will employ a sequence of stepwise transformations. A validity of each subsequent transformation will be confined to smaller and smaller energy. The first step on this path is to transform the Hubbard model (\ref{Hubbard}) to a model of magnet.  For energies well below the single particle gap $\sim U$ model (\ref{Hubbard}) is equivalent to the isotropic spin S=1/2 Heisenberg model in magnetic field which role is played by chemical potential $\mu$:
 \bea
 && H_{eff} = \sum_{{\bf r} = a_0n_x}\Big\{ J_x\Big[- \frac{1}{2}\Big(\tau^+({\bf r} + {\bf x})\tau^-({\bf r}) +H.c.\Big) + \nonumber\\
&& \tau^z({\bf r} + {\bf x})\tau^z({\bf r})\Big] + 
2\mu \tau^z({\bf r})\Big\}, \label{Heis}
 \eea
 where $\tau^a$ are Pauli matrix operators, $J_x = t_x^2/U$. $\tau^{\pm}({\bf r})$ create and destroy electron pairs located at site ${\bf r}$ and $\tau^z$ stands for the deviation local charge density from 1. This transformation valid for the attractive Hubbard model is similar to a more familiar transformation for the repulsive case where low energy degrees of freedom are described by real spin operators. One important difference is  that the exchange integral of the transverse spin components in the attractive case is ferromagnetic; this signifies that the pairs carry zero momentum. The exchange interaction in model (\ref{Heis}) can be made anisotropic by including longer range charge interactions. The repulsive ones generate an Ising-like anisotropy and as a consequence will open a spectral gap at $\mu =0$ (a commensurate CDW), but this gap is removed ones the consentration of pairs deviates from 1/2. 
The coupling between the perpendicular chains has the same Heisenberg form, but with a smaller exchange integral $J_{\perp} = t^2_{\perp}/U$.

  The next step in the derivation of the low energy action proceeds by   bosonization of model (\ref{Heis}): 
  \bea
  && \tau^{\pm}(x = na_x) = \re^{\ri\sqrt{2\pi}\Phi} + C\Big[\re^{\ri\sqrt{2\pi}(\Theta + \Phi) +2\ri k_Fx} + \nonumber\\
&& \re^{\ri\sqrt{2\pi}(-\Theta + \Phi) -2\ri k_Fx}\Big] + ...\label{plus}\\
  && \tau^z(x = na_x) = k_F/\pi + \sqrt{\frac{2}{\pi}}\p_x\Theta + \nonumber\\
&& C_z\sin(2k_F x + \sqrt{2\pi}\Theta)(-1)^n +... \label{z}
  \eea 
 where dots stand for less relevant operators and $k_F = \pi n_0$.  Here the mutually dual bosonic phase fields $\Phi$ and $\Theta$ are governed by  Hamiltonian (\ref{sG}) eplained below.  The bosonized form of the Hamiltonian is valid as yet lower energies that the range of validity of (\ref{Heis}), namely, below $J_x$. I remind the reader that $n_0$ stands for concentration of pairs. In the real YBCO where the preformed pairs  coexist with the quasiparticles a relationship between $k_F$ and the concentration of carriers is not straightforward. The magnitude of $k_F$ may be even doping independent as suggested by  the experiments \cite{xray2}. 
 
  The presence of the constant term in (\ref{z}) implies that at nonzero doping each chain experiences a Zeeman-like field with wave vectors $2\pi n/Na_0$. Therefore, when $\pi - 2k_Fa_0 = 2\pi n/N$, one of the harmonics of the periodic potential strongly couples to the staggered magnetization of the $\tau$-field (charge density wave field for actual fermions) (\ref{z}). The continuum limit Hamiltonian  valid at energies below $J_x \sim t^2_x/U$ is 
\bea
&& H_{eff} = \sum_n \Big(H_{eff,x}^{(n)} + H_{eff,y}^{(n)}\Big) - \nonumber\\
&& g_{\perp}\sum_{n,m} \cos\Big\{\sqrt{2\pi}\Big[\Phi_x^n(mNa_0) - \Phi_y^m(nNa_0)\Big]\Big\}\\
&& H_{eff,x}^n = \frac{v}{2}\int \rd x \Big\{\Big[ K^{-1}(\p_x\Theta_n)^2 + K(\p_x\Phi_n)^2\Big]  -\nonumber\\
&& \lambda\sin(\sqrt{2\pi}\Theta_n + Qx)\Big\}, \label{sG}
\eea
where $[\Phi(x'),\p_x\Theta(x)] = \ri\delta(x-x')$, $g_{\perp} \sim J_{\perp}, \lambda \sim k_F$ and $Q = \frac{\pi}{a_0}(1-2n/N)-2k_F$.  $K=1$ for the large $U$ Hubbard model, but can be renormalized by longer range interactions. The velocity is $v \sim J$. 

 At the commensurate point $Q=0$  the spectrum of each individual chain is gapped (assuming $K < 4$). The field $\Theta$ on a given chain is pinned by the periodic potential generated by the chains crossing the given one. This is a CDW ordered insulating state where the interchain coupling $J_{\perp}$ is irrelevant. Therefore  in order to have an effect this coupling  must exceed some critical value. 

Let us concentrate on the interchain coupling. The operator 
\be
\Psi = \re^{\ri\sqrt{2\pi}\Phi} \label{OP}
\ee
taking part in the interchain coupling describes a charged bosonic field of charge $2e$ (the Cooperon). As for any Lorentz invariant theory the dispersion of particles in the sine-Gordon model (\ref{sG}) has a relativistic form: $
\epsilon = \sqrt{(vq)^2 + m^2}$. 
For  $K=1$ the spectrum consists of a degenerate triplet with $m=M$ and a singlet with $m = \sqrt 3 M$, where $M \sim \lambda^{4/(4-K)}$ is determined by parameters of model (\ref{sG}). Two components of the triplet (soliton and anti-soliton) are charged $Q = \pm 2e$. Thus Cooper pairs in the insulating state are gapped excitations which gap is directly related to the CDW amplitude. They are also coherent ones since 
operator (\ref{OP}) has a non-zero matrix element between the vacuum and a single-soliton state \cite{luk}. The corresponding Green's function is (see \cite{GNT})
\be
G(\omega,q) = \frac{Z M}{(\omega + \ri 0)^2 - (vq)^2 - M^2} +..., \label{G}
\ee
where the dots stand for the incoherent part, $M$ is the soliton mass and $Z$ is a numerical factor which exact value is of no importance here. Thе coherent part of (\ref{G}) can be represented as  a correlation function of complex relativistic massive Bose field (Cooperon). Taking into account the interchain interaction in RPA I obtain the following Cooperon Green's function for coupled crossed chains:  
\begin{widetext}
\bea
G_{RPA} = \la\la \Psi_a \Psi_b^+\ra\ra = ZM\left(
\begin{array}{cc}
(\omega + \ri 0)^2 - (vq_x)^2 - M^2 & M\tilde{g_{\perp}}\\
M\tilde{g_{\perp}} & (\omega + \ri 0)^2 - (vq_y)^2 - M^2
\end{array}
\right)^{-1},
\eea
\end{widetext}
where $\tilde{g_{\perp}} = Zg_{\perp}$. The excitation spectrum of the Cooperons is 
\bea
&& \omega^2 = M^2 +q^2/2 \pm \sqrt{(M\tilde{g_{\perp}})^2 + \frac{q^4}{4}\cos^2(2\alpha)},\nonumber\\
&& \tan\alpha = q_y/q_x.
\eea
Close to critical point $\tilde{g_{\perp}}^c=M$, at $v|q| << M$, we have 
\bea
\omega^2_- = M(M-\tilde{g_{\perp}}) + v^2q^2/2 +O(q^4).
\eea
Hence at energies $<< \tilde{g_{\perp}}$ the action can be approximated by the Lorentz invariant $\Phi^4$ theory in (2+1)-dimensions: 
\bea
S = \int \frac{\rd t\rd^2x}{2ZMa_0^2} \left(|\p_{t}\Psi|^2 - \frac{v^2}{2}|\nabla\Psi|^2 - r|\Psi|^2 - u|\Psi|^4\right), \label{cont}
\eea 
where $r = M(M- \tilde{g_{\perp}})$ and $u \sim M^2$. At $\tilde{g_{\perp}}<M$ the system is an insulator, the conductivity of pairs is $\sigma \sim \exp(- \sqrt r/T)$, at $\tilde{g_{\perp}} >M$ it is a {\it supersolid}. It is remarkable that at the critical point the sheet resistivity of pairs is \cite{joe}: 
\bea
R(\omega) = (h/1.037(4e^2))\rho(\omega/T) \label{sigma}
\eea
with $\rho(0) =1$ and $h/(2e)^2 \approx 6.45 k\Omega$. 

 By construction of model (\ref{Model}) the CDW order always includes two density waves with  $(Q_0,0)$ and $(0,Q_0)$ ($Q_0 = \pi/a_0 - 2k_F$) wave vectors. Since these CDW's are  driven by the periodic potentials of the crossed chains, they are not a result  of phase transition. With increase of temperature their magnitude simply gradually diminishes from its $T=0$ value $\la \sin(\sqrt{2\pi}\Theta_{x,y})\ra \sim M^{K/2} \sim {\lambda}^{K/(4-K)}$. Since in the superconducting phase the spectral gap is closed only in at a specific point in momentum space, namely at ${\bf q} =0$,  the CDW amplitude, which magnitude is related to the average gap, in that phase does not vanish. The characteristic temperatures at which the CDW order is noticable are of the order of $M$. On the other hand, the superconducting order occurs at temperatures determined by the stiffness and hence related to the mean field value of $|\Psi|$: $T_c \sim -rv^2/u$ and hence sufficiently close to the QCP, where  $M >> T_c$, the CDW order predates the superconductivity. This conclusion agrees with the observations  in YBCO \cite{xray}. It is also important that, due to the breaking of translational invariance, the superconducting stiffness is not directly related to the density of pairs, but is non-universal being determined by parameters of model (\ref{cont}).

 To make contact with real cuprates two additional factors should be considered: quasiparticles and disorder. The presence of quasiparticles in the normal state of underdoped YBCO is well attested by the ARPES measurements done in zero magnetic field \cite{pdj}  and by the de Haas van Alphen measurements done in strong magnetic field.  The existing difficulties in figuring out an exact shape and evolution of their Fermi surface are not relevant for the present discussion.  Whatever their FS is, the quasiparticles can be easily accomodated in the present model, as it was  discussed in \cite{chub}. The action of the preformed pairs (\ref{cont}) should be augmented by the  action of quasiparticles. As I said in the beginning the thermodynamics is  dominated by the collective modes, but the normal state transport is dominated by the quasiparticles. This is due the fact that in the normal state the conductivity of the Cooperons cannot exceed certain value which order of magnitude is determined by (\ref{sigma}). The latter one is the minimal metallic conductivity and in a clean YBCO thе quasiparticles short circuit the conductivity of the Cooperons.  

 In the region of phase diagram where  CDW was observed, YBCO is a very clean compound as shown by the mobility measurements \cite{Hall} and NMR \cite{bobroff}. The inverse life time of these quasiparticles in the strong magnetic field regime is estimated as $\sim 5 meV$ \cite{lonzarich}. A disorder  introduced in CuO planes causes two effects: the $T_c$ is suppressed and $\rho(T)$ reveals $\ln T$ upturn at low temperatures. This strong logarithmic upturn has not been explained by standard localization theories. According to \cite{alloul}, the coefficient at $\ln T$ is proportional to the impurity concentration suggesting that it originates from a strong scattering on isolated centers. This brings to mind Kondo effect as a possible explanation (as it  was suggested in \cite{alloul}). However, the conventional Kondo effect where the inelastic scattering originates from localized spins  is inconsistent with the fact that the $\ln T$ upturn seems to be impervious to strong magnetic fields applied to suppress the superconductivity. I suggest instead a charge Kondo effect where the role of local spins is played by singlet local pairs created by disorder \cite{garate}. It is essential that the disorder is not sufficiently strong to localize quasiparticles (otherwise there would be no Kondo effect). Existence of delocalized quasiparticles in the relevant range of disorder and magnetic fields is guaranteed by the fact that the $\ln T$ upturn has been observed in samples whose resistance slightly above the transition was smaller than the quasiparticle  localization shreshold: $R(T_c) < 25.4 k\Omega$. 

 Let us now imagine that we are in the CDW regime where in a clean sample pair excitations would have a finite energy. However, the disorder by radically  weakening the interaction of a given pseudospin with its neighbours can create a localized two-level state  with zero energy which strongly scatters the quasiparticles. Such localized states are described as pseudospins as in model (\ref{Heis}). An isolated pseudospin is free to flip in the process of interactions with the quasiparticles (not shown in (\ref{Model}), as I said above to describe the transport one has to add quasiparticles) which causes their inelastic scattering. The Kondo Hamiltonian describing the interaction between pseudospins and   quasiparticles is given by 
\bea
&& H_K = \sum \epsilon(k)a^+_{\s}(k)a_{\s}(k) + 2\sum_{{\bf R}}\mu({\bf r})\tau^3({\bf r}) \nonumber\\
&& V^{-1}\sum_{{\bf k},{\bf k}',{\bf r}}\re^{\ri({\bf k}-{\bf k}'){\bf r}}\Big\{\Big[\gamma_{\perp}({\bf k}-{\bf k}')\tau^+({\bf r})a_{\uparrow}({\bf k})a_{\downarrow}({\bf k}') + H.c.\Big] \nonumber\\
&& + \gamma_{\parallel}\tau^3({\bf r})a^+_{\s}({\bf k})a_{\s}({\bf k}')\Big\}, 
\eea
where $a^+_{\s},a_{\s}$ are creation and annihilation operators of quasiparticles. The $d$-wave symmetry of the order parameter wave function is accomodated in the $k$-dependence of the coupling constants.  A strong contribution to electron scattering comes from the pseudospins whose chemical potential is smaller than the Kondo temperature $T_K$ (its value  depends on the coupling constants and the quasiparticle bandwidth).  Such pseudospins will dominate the resistivity provided the distribution function of $\mu(r)$ has a sharp peak at small $\mu$'s.

 In this paper I have shown that  the theory formulated in \cite{chub} brings together such features of low energy physics of the underdoped YBCO as coexistence of superconductivity and charge density wave and the $\ln T$ upturn in the resistivity. The latter effect is caused by disorder and quasiparticles. 
 It is quite possible that the model has a more general applicability since there are indications of low doping QCP in other cuprates.  

 The model postulates existence of intermediate energy scale $\tilde g_{\perp}$ (smaller than the pseudogap) serving as the upper cut-off for the theory  (\ref{cont}). In \cite{chub} it was associated with the emergence of enhanced diamagnetism; it may also be related to the emergence of CDW observed by x-ray scattering \cite{xray},\cite{xray2}.  As I have already mentioned within the present setting this order emerges gradually and is not accompanied by any thermodynamic anomalies.  The change of sign of the Hall coefficient \cite{Hall},\cite{Hall2} occuring in the the same temperature range may also originate from the quasiparticle Fermi surface reconstruction caused by the CDW. ARPES experiments on BiSCO also point towards existence of such intermediate energy scale below the pseudogap \cite{adam}. 

  I am grateful to A. Chubukov, T. M. Rice, K. Efetov and J. Tranquada for constructive criticism and interest to the work. This was supported by the Center for Emergent Superconductivity, an Energy Frontier Research Center, funded by the Office of Basic Energy Sciences (BES), Division of Materials Science and Engeneering, US Depertment of Energy, through Contract No. DE-AC02-98CH10886. 


\end{document}